\documentclass[runningheads]{llncs}

\usepackage{listings}
\usepackage{amsmath,amssymb,amsfonts}
\usepackage{algorithmic}
\usepackage{graphicx}
\usepackage{textcomp}
\usepackage{xcolor}

\usepackage{lineno}
\usepackage{mdframed}

\usepackage{caption}
\usepackage{subcaption}
\usepackage{flushend}

\def\BibTeX{{\rm B\kern-.05em{\sc i\kern-.025em b}\kern-.08em
    T\kern-.1667em\lower.7ex\hbox{E}\kern-.125emX}}

\usepackage{url}

\usepackage{breakurl}
\usepackage[breaklinks,hidelinks]{hyperref}

\usepackage{enumitem}
\usepackage[final]{microtype}

\newcommand{\URIR}{\mbox{URI-R} }

\newcommand{\URIG}{\mbox{URI-G} }

\newcommand{\URIRs}{\mbox{URI-Rs} }
\newcommand{\URIMs}{\mbox{URI-Ms} }
\newcommand{\URITs}{\mbox{URI-Ts} }

\newcommand{\URIMns}{\mbox{URI-M}}
\newcommand{\URIRns}{\mbox{URI-R}}

\newcommand{\URIMsns}{\mbox{URI-Ms}}

\newcommand{\todoHidden}[1]{\ }

\newcommand{\red}[1]{{\color{red} {#1}}}
\newcommand{\blue}[1]{{\color{blue} {#1}}}
\newcommand{\orange}[1]{{\color{orange} {#1}}}
\newcommand{\black}[1]{{\color{black} {#1}}}
\newcommand{\greenish}[1]{{\color{green!40!black} {#1}}}
\newcommand{\urimcolor}[1]{{\color{purple} {#1}}}

\makeatletter
\AtBeginDocument{%
  \def\doi#1{\url{https://doi.org/#1}}}
\makeatother

\begin{document}

\title{Aggregator Reuse and Extension\\for Richer Web Archive Interaction}

\author{Mat Kelly\orcidID{0000-0002-0236-7389}}

\authorrunning{M. Kelly}

\institute{Drexel University, Philadelphia PA 19104, USA\\
\email{mkelly@drexel.edu}\\
\url{https://matkelly.com}}

\titlerunning{Aggregator Reuse and Extension for Richer Web Archive Interaction}
\maketitle

\begin{abstract}
Memento aggregators enable users to query multiple web archives for captures of a URI in time through a single HTTP endpoint. While this one-to-many access point is useful for researchers and end-users, aggregators are in a position to provide additional functionality to end-users beyond black box style aggregation. This paper identifies the state-of-the-art of Memento aggregation, abstracts its processes, highlights shortcomings, and offers systematic enhancements. 

\end{abstract}

\section{Introduction}
\label{sec:intro}

Web archives act as a historical record of the web. The Internet Archive (IA) possesses the largest number of web archive holdings. These holdings are accessible through a set of interfaces to the Wayback Machine. Beyond IA, other web archives exhibit focused collection efforts, often providing unique captures within IA's temporal and spatial (i.e., URL \cite{rfc3986}) voids \cite{lobbe2018dead}. A common usage pattern in accessing IA's captures is to request the archive's web site at \url{archive.org}, submit a URL of interest by providing it in a text input field, then selecting a date and time from the set of available captures for that URL in the past. This pattern may differ between web archives' respective web interfaces. \mbox{Memento \cite{rfc7089}} provides the standards-based interoperable means, dynamics, syntax, and semantics for representing identifiers for archival captures (mementos) from a set of web archives. Each archive that supports the Memento Framework provides an HTTP endpoint for retrieving mementos from their respective archival holdings. Users can send a request for all captures of a URL to a variety of supporting archives through a single endpoint by an accessible tool that performs the logic of querying and combining results from multiple sources---a Memento aggregator.

\begin{figure}[h]
\centering
\fbox{
\begin{minipage}{0.75\linewidth}\scriptsize\color{gray}
\texttt{\black{t$_0$:} \blue{\{scheme \& hostname\}}/\orange{\{resource type\}}/\black{\{format\}}/\red{\{URI-R\}}}\\
\texttt{\black{t$_1$:} \blue{https://myarchive.org}/\orange{timemap}/\black{link}/\red{http://example.com}}\\
\texttt{\black{m$_0$:} \blue{\{scheme \& hostname\}}/\greenish{\{datetime\}}/\red{\{URI-R\}}}\\
\texttt{\black{m$_1$:} \blue{http://archive.md}/\greenish{20210619183508}/\red{https://icadl.net/icadl2021/}}\\
\texttt{\black{m$_2$:} \blue{https://archive.ph/eoQRZ}}
\end{minipage}
}
  \caption{An aggregator must be configured to supply parameters to an HTTP endpoint (like t$_1$), often exhibited in the form of a ``templated URI'' (t$_0$) for a URI-T as shown here. The suffixed {\color{red} red portion} represents a \URIR http://example.com as used in practice. This URI templating is replicated (m$_0$) with \URIMs (e.g., m$_1$), though a web archive need not identify its captures in this non-opaque manner (m$_2$ and m$_1$ identify the same memento).}
  \label{fig:templatedURI}
  \end{figure}

Memento aggregators typically have reference to a set of endpoints to web archives that implement the Memento Framework. An aggregator may express this through a URI ``template'' like Figure~\ref{fig:templatedURI} or as a URI with an implicit append operation of a \URIR \cite{rfc7089}. Upon receiving a request from a client with a parameterized URL (e.g., the \URIR applied to the template URI), an aggregator relays the argument received in this request as parameters for subsequent requests to each archive. When the aggregator receives a sufficient response\footnote{This criteria is implementation-specific and may be associated with a temporal threshold, memento count, etc.}, as dictated by the logic of the aggregator in-practice, the aggregator combines the results through a procedure that aligns with Memento syntax, often inclusive of temporal sorting\footnote{It is important to note here that TimeMaps do not need to be temporally sorted to be Memento compliant.}. The aggregator returns this ``aggregated'' response to the client. This description somewhat encompasses the conventional role of the aggregator. Its place as a means for users to interface with multiple web archives through a single request has the potential to be further utilized, exploited, and be more generally useful.

This paper examines the hierarchical (yet decoupled) relationship between a Memento aggregator and Memento-compliant web archives. While an aggregator and a set of archives often exhibit a static one-to-many relationship (respectively), there exists both more fundamental and more potentially complex hierarchies that may be exhibited using existing infrastructure. These exhibitions may be strategically and efficiently enhanced through consideration of this potential additional capability for the sake of enhancing the role of the aggregator in use cases for web archives. We build on existing work in defining a framework for aggregating public and private web archives \cite{kelly-jcdl2018}. Our focus will be on identifying (Section~\ref{sec:barriers}) and mitigating (Section~\ref{sec:extensions}) some outstanding issues both introduced by the framework as well as those that exist in current practice of interfacing with web archives using Memento aggregation.

\section{Background}
\label{sec:background}
The Memento Framework \cite{rfc7089} introduces the ability to perform temporal negotiation on the web by relating the current and past representations of a web page. Past representations are identified by ``URI-Ms'' and the original representation by a ``\URIRns'', per Memento. Memento also introduces a resource to associate \URIMs and \URIRs through a structured listing called a TimeMap, identified by a ``URI-T''. A web archive may return a TimeMap representing its holdings, inclusive of \URIMsns, a URI-R, URI-Ts, and a URI-G for a ``TimeGate''. A TimeGate allows a client, through HTTP request headers, to specify a datetime basis for a likewise included URI-R. This paper relates to the information retrieval and relational aspects of Memento TimeMaps and not specifically to the temporal negotiation of Memento, the latter being a feature of TimeGates. We focus on the association of past and present URIs and not the ability to resolve the closest datetime, both of which Memento provides.

\begin{figure*}[t]
\centering
  \begin{subfigure}[b]{0.49\textwidth}
  \includegraphics[width=1.0\linewidth]{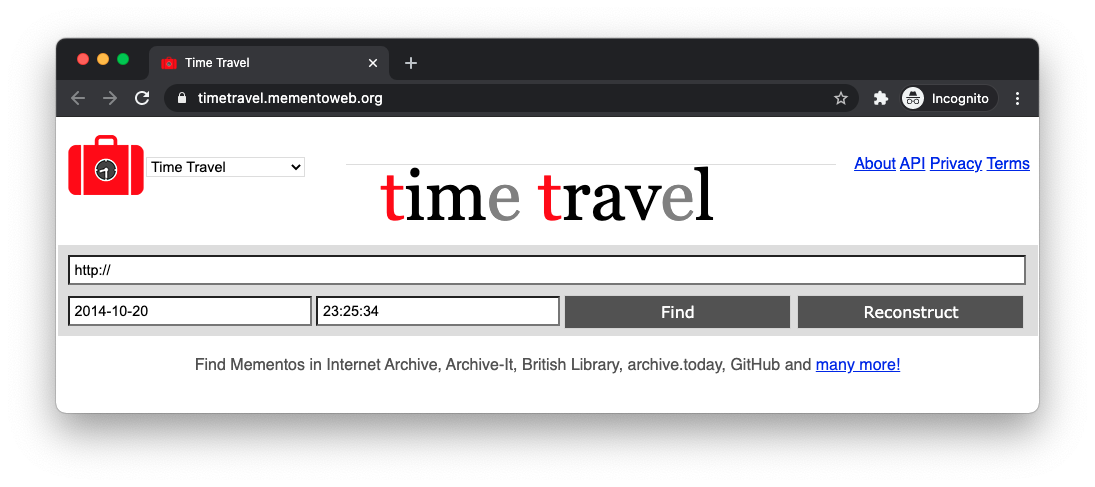}
  \caption{}
  \label{fig:timetravel}
  \end{subfigure}
  \begin{subfigure}[b]{0.49\textwidth}
  \centering
  \includegraphics[width=1.0\linewidth]{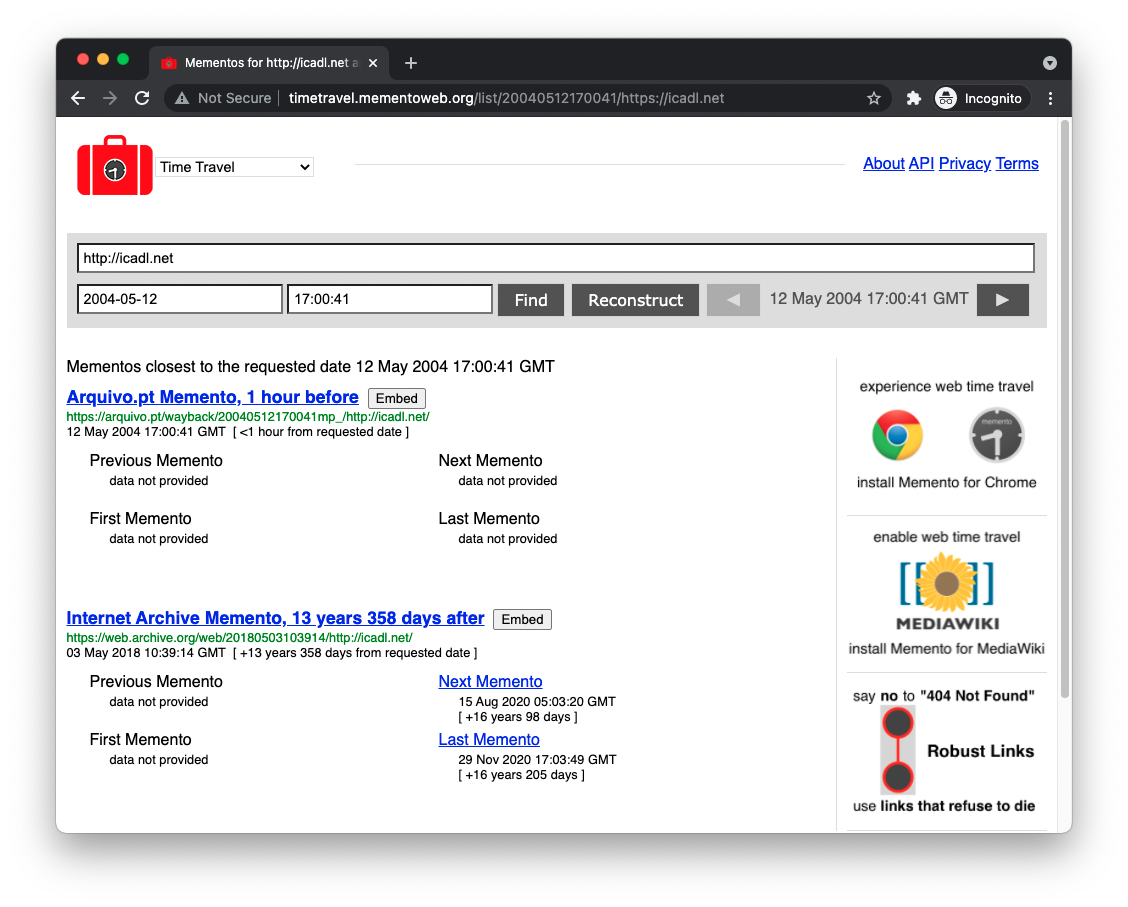}
  \caption{}
  \label{fig:timetravel_results}
  \end{subfigure}
  \caption{The ``Time Travel'' service provides a graphical, web-based endpoint to interface with LANL's Memento aggregator. After submitting a URI and date range in the interface (\ref{fig:timetravel}), the results are displayed (Figure~\ref{fig:timetravel_results}), showing the extent of the captures from a variety of pre-configured, server-defined web archives.}
  \label{fig:mementoweb}
\end{figure*}

The concept of aggregation goes beyond the Memento specification by leveraging a similar structure to TimeMaps but allowing the URIs contained within the aggregated TimeMap to identify resources at multiple archives instead of a single archive. The Research Library at Los Alamos National Laboratory (LANL) deployed the original Memento aggregator \cite{bornand,jones-memento21}, currently accessible through a web interface via the Time Travel service at \url{https://timetravel.mementoweb.org/}. This web service (Figure~\ref{fig:timetravel}) provides an HTML form field for a user to specify the URI-R and a datetime then uses temporal negotiation to query a set of archives and return links to the results (Figure~\ref{fig:timetravel_results}).

A central point of access also implies a central point of failure---if the aggregator goes down, no further aggregation may be performed, and users must again resort to querying individual web archives. In response, Alam and Nelson created MemGator \cite{memgator}, a portable, open-source, cross-platform, user-deployable Memento aggregator. This tool enables individuals to no longer solely rely on a single web-accessible aggregator but also configure, use, and potentially deploy their own. Also, unlike Time Travel, a user has the ability to control \textit{which} web archives are queried for mementos. This newfound ability provided the accessibility of the aggregation capability to be further explored by researchers. 

Memento is an extension to the Hypertext Transfer Protocol (HTTP). HTTP is a stateless, client-server based protocol on which the web is built. In the context of Memento, a client provides an HTTP request for a TimeMap of a URI in the past, often by appending a URI-R to a templated endpoint (Figure~\ref{fig:templatedURI}). Both the identifiers for a TimeMap and a memento are returned with corresponding \mbox{Link \cite{rfc8288}} HTTP response headers giving additional context to the representation. A user (e.g., person) will typically act as a client through a user-agent (e.g., web browser, cURL\footnote{\url{https://curl.se/}}) and may send an HTTP request to a Memento aggregator with the expectation of receiving an HTTP response. The aggregator, in-turn, acts as a client to the web archives, relaying the request for the URI-R in the past and expects HTTP responses. This use case of a Memento aggregator playing the role of a server and a client is abridged in Section~\ref{sec:mink}.

\section{Related Work}
\label{sec:relatedWork}
Most research involving Memento aggregation relates to usage of the aggregator rather than enhancement of the aggregation process. In the same way that prior to MemGator, researchers would state ``we requested URIs from the Time Travel Service'', this statement was transformed to ``we used MemGator to request URIs'', indicative that it was useful for researchers to utilize their own aggregator instance \cite{nwala-jcdl2017,kelly-jcdl2017,lulwah-tois}. A facet of this use case is the ability for researchers to customize the set of web archives to be used as the basis for querying, which is performed prior to running MemGator by modifying a configuration file\footnote{An aside: researchers that need to control the process do so either through manipulation of their internal software (LANL experimenting with Time Travel \cite{bornand}) or those outside of LANL utilizing MemGator.}. This paper examines the aggregation process beyond accessing an aggregator and does so at a more abstract level than the ability to customize the archival sources.

\subsection{Using Aggregators Beyond End-User Aggregation}

As MemGator is free and open-source software (cf. Time Travel), many research endeavors on evolving the aggregation process have centered around enhancing its development beyond the limited endpoint-based Time Travel ecosystem. While the set of archives to be aggregated is static, both in accessing the Time Travel service as well as a deployed MemGator instance, other standards-based mechanisms like HTTP Prefer \cite{rfc7240} provide a means of allowing a client to specify the set of archives aggregated to an ``enhanced'' aggregator---in this case, an extended version of MemGator \cite{wadl-prefer}. This approach \cite{wadl-prefer} entailed encoding the set of archives that normally reside in a server-side configuration file to be customizable at \textit{query} time. The specification of custom archival sources utilizes the ``Prefer'' HTTP request header with a value being the self-describing, base-64 encoded JSON representing the aggregator's configuration of endpoints. A prototypical extension of MemGator referenced by the authors required the aggregator to read the HTTP request header and respond accordingly at runtime to request captures only from the archives specified by the client. 

\begin{figure*}[t]
\begin{subfigure}[b]{0.47\textwidth}
\centering
  \includegraphics[width=1.0\linewidth]{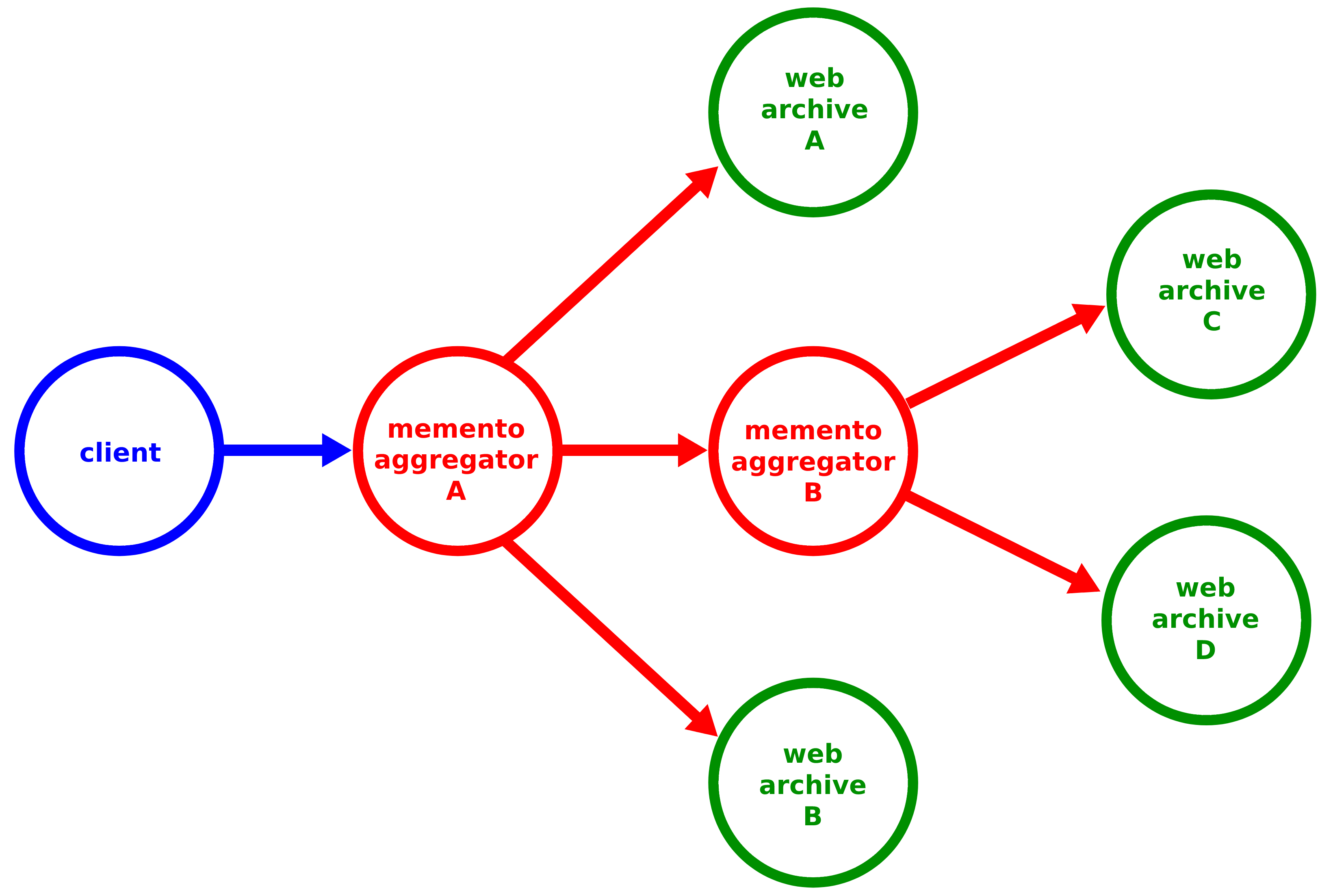}
  \caption{}
  \label{fig:hierarchy}
  \end{subfigure}
\begin{subfigure}[b]{0.47\textwidth}
  \centering
  \includegraphics[width=0.95\linewidth]{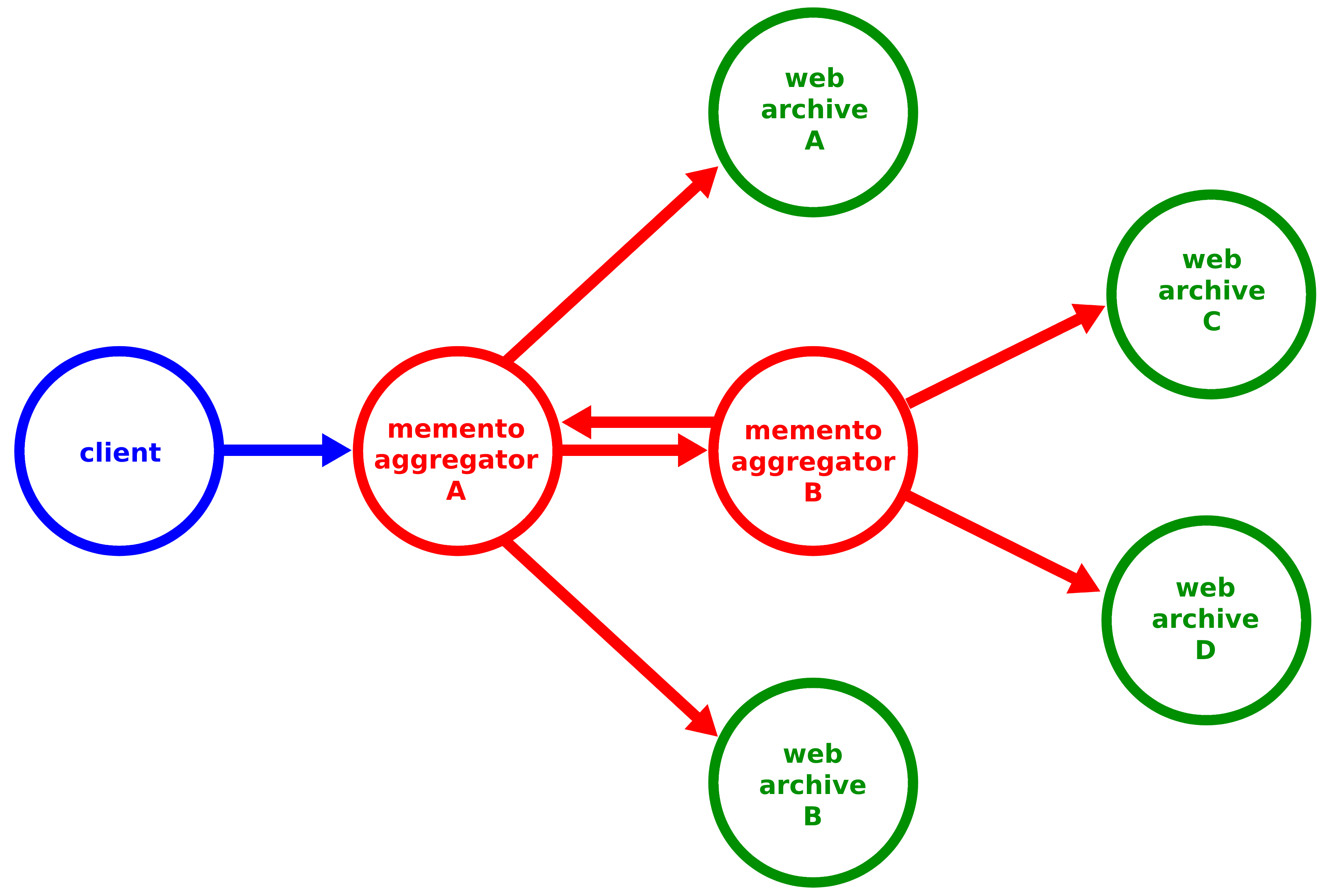}
  \caption{}
  \label{fig:infiniteloop}
  \end{subfigure}
  \caption{An aggregator is configured to query HTTP endpoints (Figure~\ref{fig:hierarchy}), which are typically from web archives, but could equally be configured to be to other aggregators causing an ``aggregator chaining'' effect (Section~\ref{sec:s2}). Aggregators are agnostic of whether their requester is a client, script, or aggregator itself (Figure~\ref{fig:infiniteloop}) and thus may send a request that ultimately resolves to a requester causing an infinite loop.}
  \label{fig:combined}
\end{figure*}

\subsection{Abstractions from Other Domains}

The process of HTTP requests as recursively applied through an aggregator subsequently querying additional sources resembles a graph structure, typically reduced to a tree in the conventional case (Section~\ref{sec:s1}). As this work reiterates the potential for an aggregator querying an aggregator \cite{kelly-jcdl2018}, the scenario arises of graph-style cycles (Figure~\ref{fig:combined}) that must be mitigated. Additionally, we may encounter redundancies in this ``chaining'' process (Figure~\ref{fig:dupesrcs}) where aggregators down the request chain are configured to query identical, previously queried archives with the same parameters. The similarity of this problem resembles a singly linked list wherein a child does not know the capacity of its parent and is in adherence of HTTP being stateless. Here, an origin node is aware of that to which it links but a node is likely not aware of the linkages from its parent, to which the node itself is one. 

\subsection{Aggregation Optimization}
\label{sec:aggregationOptimization}

The process of aggregation can be complex \cite{mln-hvds-sage-2018}, both in programmatic logic to accomplish it as well as largely so in the temporal, spatial, and computational requirements. In conventional practice (Section~\ref{sec:s1}), upon receiving a request, an aggregator will then send a request to each web archive, as defined by the endpoints in the aggregator's configuration. The process of sending these requests can typically be performed asynchronously \cite{memgator}, as the response time from a particular archive may be affected by a variety of factors including its infrastructure capabilities, the quantity of its holdings, the temporal spread of its holdings, etc.

Different web archives inherently possess a different set of archival holdings\footnote{We distinguish ``archival holdings'' from mementos in that the latter implies compliance with the Memento Framework.}. For example, an archive may only collect web pages within a limited set of ccTLDs \cite{rfc1591} like \texttt{.ac.uk} and \texttt{.gov.uk} for academic and government websites in the United Kingdom (respectively). Repeated requests for TimeMaps from web archives that consistently have no mementos for a structured type of URI produce inefficiencies that are exacerbated when aggregated and affect the aggregation process. AlSum et al. \cite{alsum-profiling} generated profiles to identify the distribution of URIs across archives and the effect on recall by both including and excluding IA from the aggregated results. MementoMap \cite{alam-mementomap} provided an approach to remedy this issue with the cooperation of a web archive. By an archive supplying indexes of its holdings, a ``map'' can be created to abstractly represent (using wildcards) the extent of the holdings for specific URI patterns. This may be abstracted to the level of TLD (e.g., the extent of the holdings within the \texttt{.uk} TLD) down to the specificity of the quantity of holdings within a specific path of the URI. MementoMaps also provide a format to represent this extent both on the level of URI-R and URI-M. Through the cooperation of one such scoped archive, the Portuguese Web Archive, Alam et al. \cite{alam-mementomap} were able to demonstrate the increase in efficiency of selectively sending requests to a subset of archives informed by their respective holdings. This work leveraged MemGator. Aturban et al. \cite{aturban-tpdl21}, through a longitudinal study on the web archives themselves, identified the disappearance of the base URI of an archive, further highlighting the need for an aggregator to be updated to ensure resolution as archives change their hostnames.

In related work, Bornand et al. \cite{bornand} consulted logs from the aggregator created by the Time Travel service (the authors are from LANL) to create classifiers to effectively route queries rather than relying on a web archive to provide a profile. They analyzed over 1.2 million URI-Rs from the aggregator's cache (with over 239,000 URI-Ms) to identify a point-of-compromise for optimizing the requests sent to an archive based on the true and false positive rate as informed by prior requests.

Part of this work entails enabling the user to have more extensive interaction with web archives using Memento. This is frequently enabled through the use of browser extensions \cite{mink,mementofox} and dedicated applications \cite{tweedy-memento,mobilemink,mementoForChromeWSDLBlag}. Mink\footnote{\url{https://github.com/machawk1/mink}} is an extension for the Chrome web browser that allows a user to extend the context of the web page they are currently viewing to be used as the basis of a request to a Memento aggregator. Some preliminary efforts have been performed to provide further user control over archival selection from the web browser using the extension, but have not been formalized nor deployed in the primary extension. 
 Doing so entails either the approach of requiring an enhanced aggregator that receives a request to adapt their set of archives queried at runtime based on the user's request (a server-side approach) or for Mink to filter the results on the client after the aggregator returns the results. In the latter, client-side approach, the logic of aggregation becomes the responsibility of the extension when an aggregator does not comply with sending requests to archives outside of its base configuration.

\section{Base Querying Models}
Per Section~\ref{sec:relatedWork}, Memento aggregators are often configured to be used as a web service; in the case of MemGator, specifying a list of archives, timeouts, etc.; and ``used'' by querying the aggregator's HTTP endpoints with the URI as a parameter. In this Section we define aggregator ``querying models'' for further discussion.

\subsection{Proxy-style Querying (S\textsubscript{0})}
An aggregator may be configured to query a single web archive. This is typically not exhibited because of redundancy (i.e., the user would normally just send the request to the archive directly), but serves as a base case for the querying models for further discussion. Here, the ``aggregator'' acts as a simple relay or proxy between the client and the web archive. This might potentially be useful for specifying a configuration to the aggregator beyond what can be expressed with a request to URI\footnote{Tools like cURL can also specify timeouts as command-line flags, but this moves the responsibility to the client}, e.g., timeouts for a response.

\begin{figure*}
  \begin{mdframed}[font=\scriptsize]\internallinenumbers
\begin{flushleft}{\ttfamily\color{black}
\$ curl https://memgator.example/timemap/link/https://icadl.net/\\
\ \\
{<\orange{https://icadl.net}>; rel="original",}\\
<\greenish{https://memgator.example/timemap/link/https://icadl.net}>; rel="self"; type="application/link-format",\\
<\urimcolor{https://web.archive.org/web/20180503103914/http://icadl.net/}>; rel="first memento"; datetime="Thu, 03 May 2018 10:39:14 GMT",\\
<\urimcolor{https://web.archive.org/web/20200815050320/https://icadl.net/}>; rel="memento"; datetime="Sat, 15 Aug 2020 05:03:20 GMT",\\
<\urimcolor{https://web.archive.org/web/20200826164340/https://icadl.net/}>; rel="memento"; datetime="Wed, 26 Aug 2020 16:43:40 GMT",\\
<\urimcolor{https://web.archive.org/web/20201101023226/https://icadl.net/}>; rel="memento"; datetime="Sun, 01 Nov 2020 02:32:26 GMT",\\
<\urimcolor{http://web.archive.org/web/20220602205625/https://icadl.net/}>; rel="last memento"; datetime="Thu, 02 Jun 2022 20:56:25 GMT",\\
<\greenish{https://memgator.example/timemap/link/https://icadl.net}>; rel="timemap"; type="application/link-format",\\
<\greenish{https://memgator.example/timemap/json/https://icadl.net}>; rel="timemap"; type="application/json",\\
<\greenish{https://memgator.example/timemap/cdxj/https://icadl.net}>; rel="timemap"; type="application/cdxj+ors",\\
<\blue{https://memgator.example/timegate/https://icadl.net}>; rel="timegate"}
\end{flushleft}
\end{mdframed}

  \caption{A typical use case for a Memento aggregator is for a user to specify a URL and receive a TimeMap representing a list of identifiers (URI-Ms) in the past---S\textsubscript{1}. Shown here is a Link \cite{rfc8288} formatted aggregated TimeMap from MemGator containing a \URIR (line 3 in \orange{orange}), \URITs (lines 4, 16--21 in {\color{green!40!black}green}), \URIMs (lines 6--15 in \urimcolor{purple}) and a \URIG (line 22 in {\color{blue}blue}).}
  \label{fig:S1}
  \end{figure*}

\subsection{Conventional Querying (S\textsubscript{1})}
\label{sec:s1}
Typical aggregator usage entails a client sending a request to an aggregator that then queries multiple web archives, aggregates the responses, and returns this response to the client (Figure~\ref{fig:S1}). The internal logic of the aggregator is not necessarily as relevant in defining this model but is critical for an aggregator's operation. For example, an aggregator may pipeline the requests for more efficient querying. An aggregator also might require archives to respond within a time threshold and ``short-circuit'' the response to disregard archives that do not respond in time. The abbreviated set of results could then be aggregated based on the subset archives that have responded up to that point in time. Some of these aspects are discussed further in Section~\ref{sec:extensions}.

\subsection{Aggregator Chaining (S\textsubscript{2})}
\label{sec:s2}
A Memento aggregator may successfully query any endpoint that is Memento compliant. The response from an aggregator is itself also typically Memento compliant. This begets the possibility that what is typically considered a ``web archive'' configured as an endpoint to query by an aggregator may be an aggregator itself, i.e., an aggregator querying an aggregator (Figure~\ref{fig:hierarchy}). One reason this is not typically exhibited is because the set of archives that are queried are (in practice) manually validated before being put in-place in the configuration. In the case of the Time Travel service, there is no indication that an aggregator is queried by the basis aggregator handling the initial response. For MemGator, however, the set of endpoints is user-configurable, and thus this valid scenario may arise and has implications. The merits of ``aggregator chaining'' were discussed in the seminal work introducing the concept \cite{kelly-jcdl2018}, but did not go into detail or highlight some problems that may occur. We reiterate and address these in Section~\ref{sec:barriers}.

As above, an aggregator may plausibly query a second aggregator. More fundamentally, and problematically, an aggregator can specify itself in its own definition of sources to query. This can be mitigated by the aforementioned manual validation, but the more scalable and programmatic approach might be accomplished through short-circuiting conditional logic in the querying function, i.e., preventing an aggregation web service from sending a request to itself and causing an infinite loop (Figure~\ref{fig:infiniteloop}). Doing so in the self-referencing case is straight-forward but through the indirection introduced through aggregator, an ``aggregator-in-the-middle'' prevents this logic from being enforced, as a request from a secondary aggregator would be handled as if from any other client. We discuss this problem further in Section~\ref{sec:cycles}.

\section{Core Features}
\label{sec:coreFeatures}

In this paper we define approaches to extend the capability of the aggregator abstraction without regard to implementation. This brief but important Section defines the empirical assumptions and expectations currently exhibited by an aggregator. These premises of an aggregator set forth the foundational base cases of expectations of an implementation. We build on these assumptions in Section~\ref{sec:extensions}.

\begin{description}
\item[\textbf{Expectation 1}] An aggregator must treat web requests received as clients and the requests it sends to archival sources as agnostic of the dynamics of the receiver. 
\item[Expectation 2] An aggregator must treat clients' requests equally, regardless of whether a requestor is a user-agent, a script, or an aggregator itself.
\item[Expectation 3] An aggregator is unaware of whether its own configuration incurs any sources queries of its parent.
\item[Expectation 4] An aggregator must treat clients as stateless and return results from its queries sources.
\end{description}

\section{Existing Problematic Scenarios}
\label{sec:barriers}
What might be deemed as ``mis-''configuration of a Memento aggregator may only be exhibited and discoverable upon execution of a request for aggregation. Typical approaches for including a web archive as an aggregation source are (1) the popularity of the archive itself to merit inclusion, (2) manual discovery by those responsible for configuring the aggregator, or (3) efforts toward publicity on the part of the archive itself to make those responsible for the archive's existence and Memento compliance. There is no established process for an archive to declare the availability of its holdings in an effort to be included in a publicly accessible aggregator \cite{rosenthal-aggregators,rosenthal-mementomarketplace}. Web archives with restricted holdings may be unsuitable to aggregate for reason of privacy of the holdings \cite{kelly-jcdl2018} or the requirement to limit accessibility beyond the conventional public scope. For example, the UK Web Archive requires a client to be physically on-site to access some of its holdings, otherwise returning an HTTP 451 (Unavailable For Legal Reason) \cite{rfc7725} status code.

Aggregators like the Time Travel service also supply TimeGate functionality, allowing for temporal negotiation (per Section~\ref{sec:background}), which is outside of this paper's scope. As temporal negotiation requires an index for efficient selection (required for scale cf. query time indexing), an aggregator would need to retain the extent of the captures on a \URIR basis from their set of sources. As this is dynamic due to the availability of various archives' web services, the non-static nature of the set of mementos in an archive, etc., a heuristic-based approach or some form of \mbox{caching \cite{bornand}} might suffice for ``good enough'' temporal negotiation. For optimal precision of the representation of sources' holdings, runtime querying of said sources' respective indexes produces a more representative result. Thus, the abstraction of a TimeGate service being co-located with an aggregator would still succumb to the effects described in this Section. The remainder of this Section describes three effects that can plague current aggregation instances: aggregation cycles (Section~\ref{sec:cycles_intro}), self-reference (Section~\ref{sec:selfReference}), and source redundancy (Section~\ref{sec:redundantSources}).

\subsection{When a Tree Becomes a Graph}
\label{sec:cycles_intro}

As an extension of S\textsubscript{2} in Section~\ref{sec:s2}, an aggregator (A) requesting captures from a second aggregator (B) may cause a cycle if the latter aggregator is configured to query aggregator A. This can be mitigated using a few approaches, one of which we describe in Section~\ref{sec:cycles}. Figure~\ref{fig:infiniteloop} illustrates an abstract scenario where this might occur with user-configurable Memento aggregators.

\subsection{Self-Reference}
\label{sec:selfReference}
A simpler example of the abstraction where an aggregator, through the request chain, is requested to respond to a request that it initiated is exhibited in an aggregator's own endpoints being within its configuration. A web service might be naive of the URI to which it is accessible, blindly sending responses after consuming and processing the parameters in the requests received. Likewise, the solution described in Section~\ref{sec:cycles} would prevent this from occurring.

\subsection{Duplication of Sources}
\label{sec:redundantSources}
\begin{figure}[h]
\centering
  \includegraphics[width=1.0\linewidth]{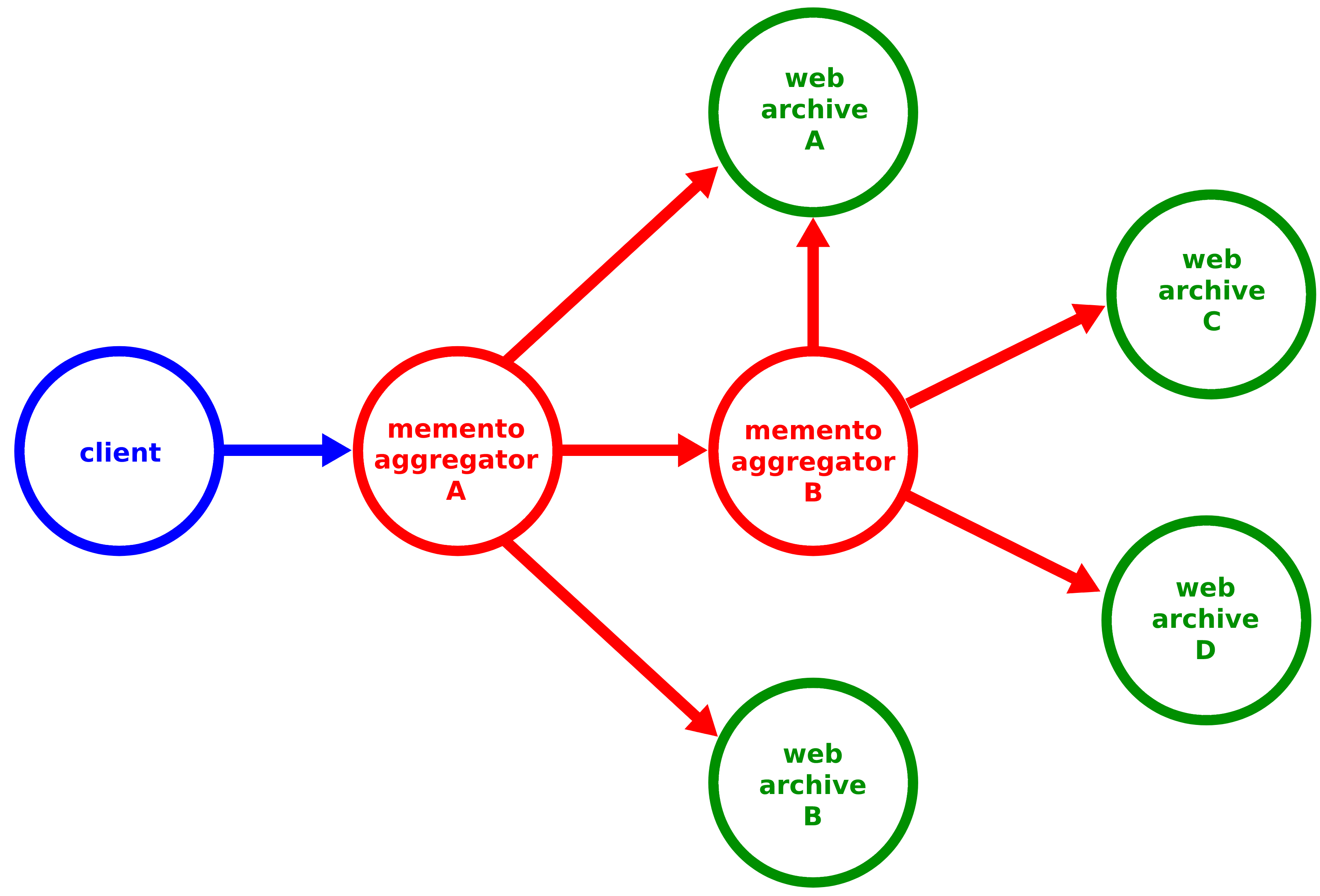}
  \caption{An aggregator (A) configured to request captures from a set of sources $\{S\}$ inclusive of a second aggregator (B) can result with B redundantly querying one of A's sources, i.e., $| S_A \cap S_B | 
  \geq 1$.}
  \label{fig:dupesrcs}
\end{figure}

The combination of aggregators being user-configurable and the potential for aggregators to query aggregators may result in duplication of results. For example, in Figure~\ref{fig:dupesrcs}, aggregator A queries web archive A, web archive B, and aggregator B. Aggregator B queries web archive A, web archive C, and web archive D. It could be useful for the clients of aggregator A to obtain the results from aggregator B, for instance, aggregator B may be privy to access restrictive web archives C and D. However, the results returned from aggregator B from web archive A will likely be redundant of those requested from aggregator A. Thus, the results may need to be deduplicated. This characteristic may also exist outside of aggregation. For instance, aggregators currently configured to request mementos from archive.org and archive-it.org (both hosted by Internet Archive) will often receive \URIMs from each archive with precisely the same 14-digit time stamp represented in the \URIMns. While it is possible that two services have unique captures (based on the tools used), this requires dereferencing the \URIMsns, which is out of the scope of this paper that focuses on TimeMaps.

\section{Newfound Capabilities}
\label{sec:extensions}
In this paper we emphasize the contribution of the untapped functional potential of a Memento aggregator beyond simple aggregation. Section~\ref{sec:coreFeatures} outlined the fundamental expectations of an aggregator that are exhibited and must be maintained as core functions. While the logic itself of strategically querying the set of archives with which an aggregator is configured has been explored in other works using profiles or machine-learning (Section~\ref{sec:aggregationOptimization}), these do not consider the breadth of potential improvements like enabling the client to have further control of the aggregation beyond URI (e.g., using HTTP Prefer \cite{wadl-prefer}), efficiency in returning partial results through HTTP endpoints, and mitigation of a non-curated set of archival sources.

\subsection{User-defined Set of Archives}
HTTP provides a standardized means \cite{wadl-prefer} for enabling the end-user (one querying an aggregator through HTTP) to specify the archival sources for aggregation -- the HTTP Prefer request header \cite{rfc7240}. The value for this header may include an encoded, modified version of the JSON data that is typically used to configure MemGator and contain custom values and transporting through the header. The expectation of an enhanced aggregator is that it will be required to decode this JSON and at its discretion, use that as the basis for the set of archives to query. Some nuances to this approach that have not been explored are (for example) whether the configuration can and should be applied to all users, the rules that should restrict which clients should be authorized to affect this change in the aggregator's operation, and how to further express the semantics to the extent to which the preference was applied (beyond supplying the Preference-Applied response header).

\subsection{Cycle Detection}
\label{sec:cycles}
In Section~\ref{sec:cycles_intro}, we introduced the potential for a cycle to occur when Memento aggregators are user-configurable and oblivious to the sources subsequently queried by aggregators further in the request chain. Approaches at mitigating cycles admittedly require the notion of HTTP being stateless to be violated. For instance, including a nonce or unique value to the request and propagating that to the sources queried (whether a web archive or aggregator), and likewise reading this value would allow the process to be short-circuited and provide a requestor some indication that the requestee was a requestor earlier in the hierarchical chain. 

\begin{figure*}[t]
\centering
\frame{\includegraphics[width=0.325\linewidth]{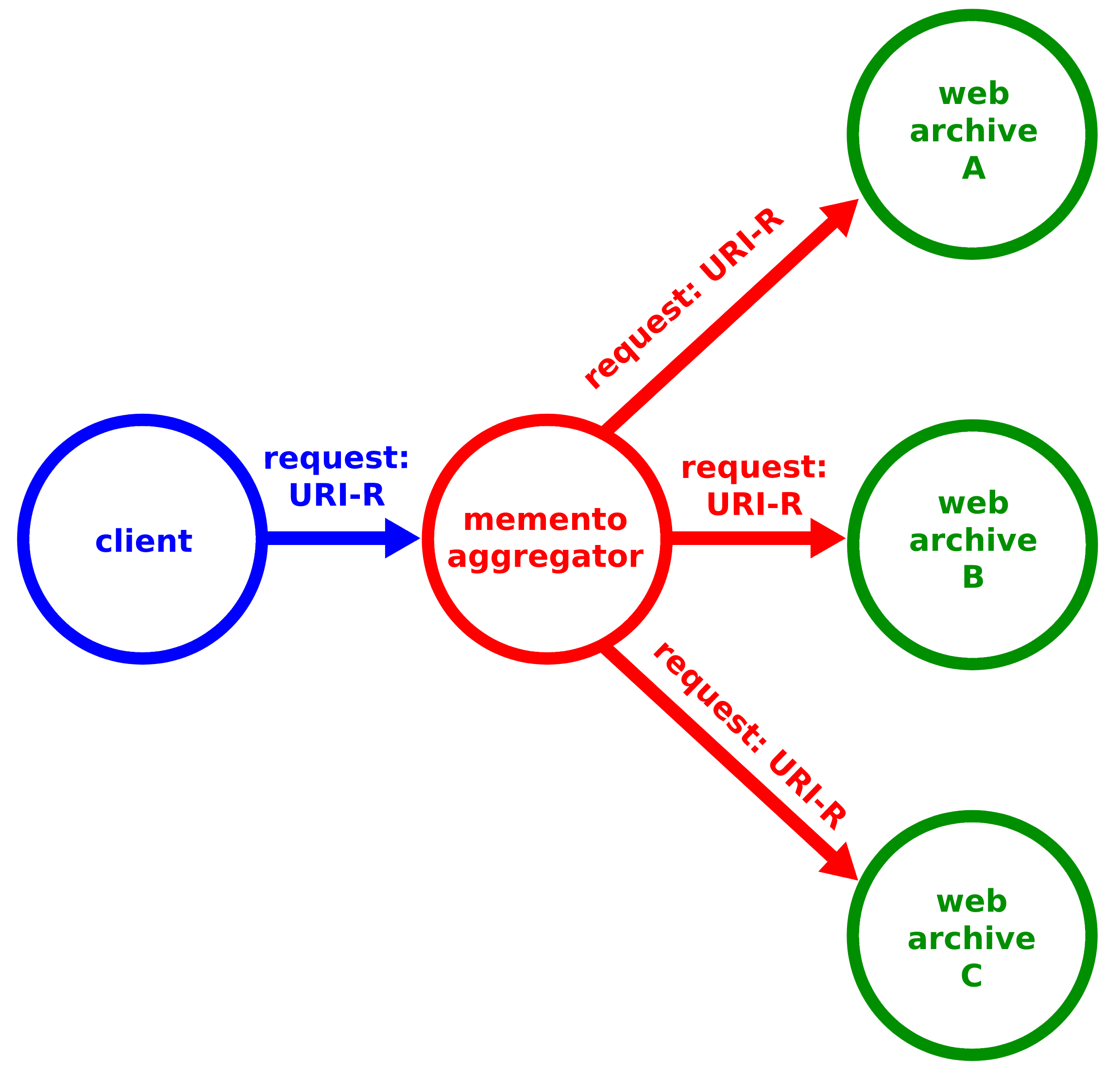}}\ 
\frame{\includegraphics[width=0.325\linewidth]{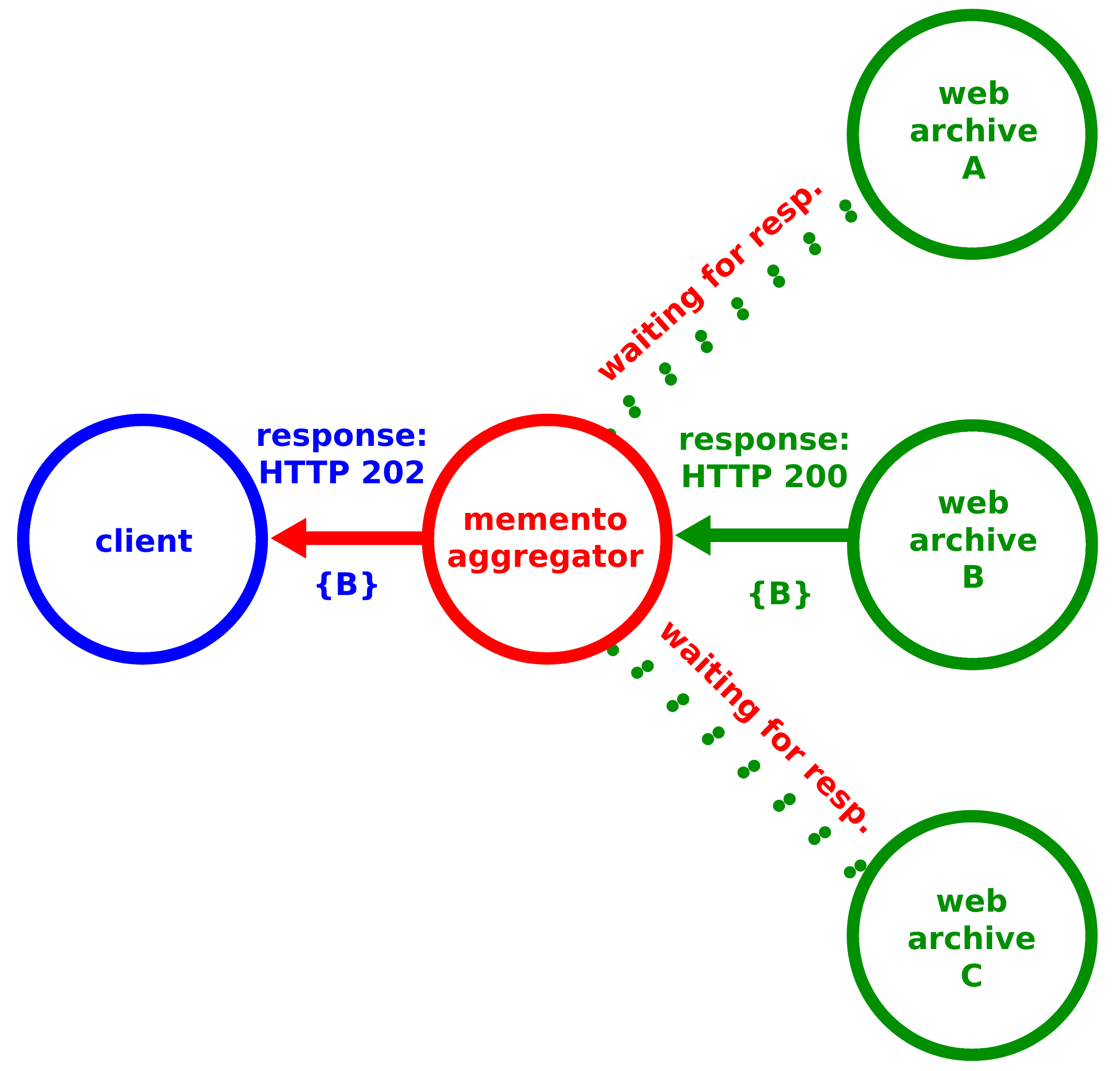}}\ 
\frame{\includegraphics[width=0.325\linewidth]{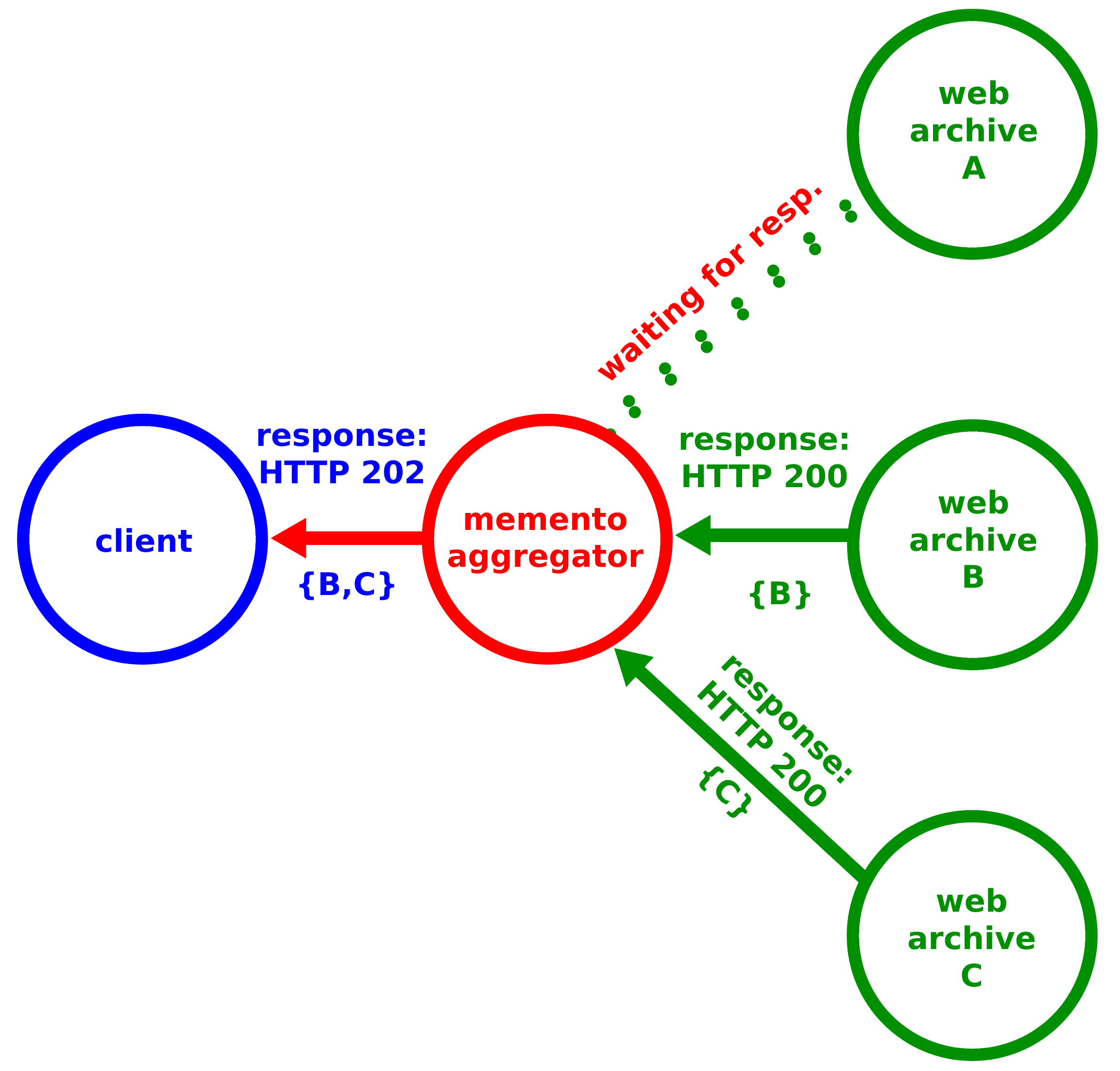}}
\caption{Rather than an aggregator waiting for the slowest archival source to respond, the response can be progressively built based on the data received thus far. This response may be served to a client as a preliminary response as indicated by HTTP 202.}
\label{fig:progressive}
\end{figure*}

\subsection{Preliminary Results Streaming}
\label{sec:streaming}
HTTP provides an often unused but standardized mechanism for a server to convey that a request is still processing (HTTP 202 status code) and that a client should wait and check back later \cite{rfc7231}, often at some indicated amount of time. In the context of Memento aggregation, web archives or other archival sources (e.g., other aggregators per Section~\ref{sec:s2}), a set of sources from which resources are requested likely returns results in respectively varying amounts of time. This can create a bottleneck while the aggregation service waits for the slowest endpoint to respond but can be optimized by progressively building the result (Figure~\ref{fig:progressive}). MemGator, for instance, merges TimeMaps as they arrive from the requesting aggregator and provide timeouts that can be specified by the user (i.e., the ``user'' that is executing the MemGator binary -- not one making the HTTP request).
  
 An important precondition for optimizing aggregators' processing through streaming is the recognition that Memento does not guarantee nor enforce internal temporal order of the identifiers in TimeMaps. When progressively merging TimeMaps from a partial set of sources requested, the merging process can be performed asynchronously relative to responses being received or more simply, not at all. For an aggregator to wait until all web archives have responded (which may never occur in the case of transient errors at an archive) is temporally inefficient. However, an incomplete (i.e., containing results only from a subset of archives), partially sorted, or unsorted aggregated TimeMap being returned to an end-user while an aggregator continues to wait can help to inform the end-user of the degree of success thus far. This may be potentially useful in cases where the results of the archives referenced in the aggregated TimeMap are explicit (e.g., through included metadata) instead of needing to be inferred (e.g., zero \URIMs from an archive \textit{might} mean no captures). This latter point can be helpful to end-users in making an informed decision to prematurely close the request if the results from an archive, as expressed in the partially aggregated TimeMap, are not to their expectations.

While the ability to return a TimeMap containing results from a subset of archives from which TimeMaps were requested may be useful and more efficient, the temporal burden for an aggregator to sort results is relatively less expensive, as it can be performed asynchronously and progressively. Despite this, partial, unsorted, concatenated TimeMaps returned using either a mechanism of streaming or through the HTTP 202 mechanism allows results, even if intermediate, to be immediately used rather than waiting on a likely unrevealed (to the end-user) set of conditions that are used prior to the response being returned.

\subsection{Rescoping the Aggregator for Client-side Execution}
\label{sec:mink}
In Section~\ref{sec:background}, we alluded to the propagation model, which may itself become recursive, of a client querying an aggregator that then similarly becomes the client through propagation of parameters. With Memento, a user-agent conventionally represents a client, transforming the request to the appropriate format (e.g., HTTP headers) as expected by a server (e.g., an aggregator).

From the client's perspective, the set of archives that an aggregator queried is not typically revealed. For example, if a client sends a request to an aggregator for \texttt{icadl.net} and receives back a TimeMap containing URI-Ms (Figure~\ref{fig:S1}), the set of archives represented by the \URIMs \textit{might} be representative of the entirety of the set, but that fact is not explicitly conveyed. It is likely and common, because of archival scoping and based on the \URIR provided, that archives within the set queried possessed no mementos for the \URIR and thus are not represented. It is wasteful and temporally inefficient to send requests to archives that possess no captures for a \URIR \cite{kelly-jcdl2018}. A priori knowledge as established by profiling archives of their holdings \cite{alam2016web} or more specifically MementoMap \cite{alam-mementomap}, helps to mitigate this problem. These advancements allow the set of archives to be strategically defined so requests for \URIRs that are unlikely to be in an archives' respective holdings are not requested. However, MementoMap requires archival cooperation and is not foolproof if the index of the captures \cite{bornand} is not updated to be representative of newly collected captures. It is also heuristic-based, so has false positive built in, i.e., likelihoods may result in no URI-Ms being returned in the TimeMap from an archive that was queried, despite their profile stating that they have captures.

\section{Discussion and Future Work}
\label{sec:futureWork}

Implicit to this work is the continuous effort to enable the end-user, for which aggregators are typically deployed, to be able to be more specific about that which they would like aggregated. As described in Section~\ref{sec:aggregationOptimization}, allowing for this degree of interaction with a web service will likely have ramifications to efficiency, for example, caching mechanism may not be beneficial if archival sources vary with each request. For the Time Travel service, this might be moot, as the set of archives queried is controlled server-side. For open-source aggregators, however, which have the potential for extended capability, this process can be further optimized and explored.

There is also the notion of functional cohesion, that is, a service should ideally do one job and do it well. This cohesion is already violated in practice with the addition of TimeGate functionality being co-located with TimeMap querying (i.e., aggregation) endpoints. We hope to see further work done in investigating use cases for both the end-user querying aggregators, researchers deploying their own aggregators, and the functions and processes inherent to the aggregation procedure to enhance the capability to make the aggregation concept generally more usable.

\section{Conclusion}
This paper focused on the aspect of Memento aggregation. We identified the state-of-the-art in pure server-side aggregators (Time Travel) and user-deployable aggregators (MemGator). Through an aggregator being user-configurable and -deployable, which has proven useful to researchers, other potential issues may arise based solely on the current functionality of an aggregator. We proposed further functional extensions to the internal aggregation process. 

From the perspective of a web service where a client sends an HTTP request to an endpoint, the aspects of this work may not much matter. However, the capacity of aggregators in the status quo still contains untapped potential capability beyond that the typical use case (S\textsubscript{1}). By enumerating these potential concerns with a user-controlled Memento aggregator, the ultimate goal of enabling a client to have more expression and preference in the process of aggregating web archives will hopefully be improved.

\subsubsection{Acknowledgements} For initial discussions on aggregator chaining and potential pitfalls, we would like to thank Chuck Cartledge, Sawood Alam, Michael Nelson, and Michele Weigle.

\pagebreak

\bibliographystyle{splncs04}
\bibliography{cyclic} 

\end{document}